\documentclass[letterpaper, 10 pt, conference]{IEEEtran2}
\IEEEoverridecommandlockouts

\usepackage{cite}
\usepackage{amsmath,amssymb,amsfonts}
\usepackage[utf8]{inputenc}
\usepackage{algorithmic}
\usepackage{graphicx}
\usepackage{comment}
\usepackage{textcomp}
\usepackage{hyperref}
\usepackage[dvipsnames]{xcolor}
\usepackage{multirow}
\usepackage{soul}
\usepackage{lipsum}
\usepackage{placeins}
\usepackage{tikz}
\usetikzlibrary{%
  arrows,%
  automata,%
  shapes.misc,
  shapes.arrows,%
  chains,%
  calc,%
  matrix,%
  positioning,
  scopes,%
  decorations.pathmorphing,
  shadows%
}

\usepackage{listings}
\usepackage{caption,setspace}
\DeclareCaptionFormat{listing}{\rule{\dimexpr0.9\columnwidth+17pt\relax}{0.4pt}\par\vskip1pt#1#2#3}
\newcounter{code}



\DeclareMathVersion{sans}
\SetSymbolFont{operators}{sans}{OT1}{cmbr}{m}{n}
\SetSymbolFont{letters}  {sans}{OML}{cmbrm}{m}{it}
\SetSymbolFont{symbols}  {sans}{OMS}{cmbrs}{m}{n}

\lstnewenvironment{sflisting}[1][]
  {\lstset{#1}\mathversion{sans}}{}

\usepackage[normalem]{ulem}

\usetikzlibrary{%
  arrows,%
  shapes.misc,
  shapes.arrows,%
  chains,%
  matrix,%
  positioning,
  scopes,%
  decorations.pathmorphing,
  shadows%
}

\usepackage[inkscapearea=page]{svg}

\begin{document}

\title{SROS2: Usable Cyber Security Tools for ROS 2}

\author{Víctor Mayoral-Vilches$^{1}$$^{,2}$, Ruffin White$^{3}$$^{,4}$$^{,5}$,  Gianluca Caiazza$^{4}$$^{,5}$,  Mikael Arguedas$^{6}$
\thanks{This material is based upon work funded by Alias Robotics and supported by the ROS4DEV project and also by the Centro para el Desarrollo Tecnológico Industrial (CDTI) under grants SEGRES (grant EXP 00131359 / MIG-20201041) and ROBOTCYSEC projects.  This work has been partially supported by the project VIR2EM - VIrtualization and Remotization for Resilient and Efficient Manufacturing” – POR FESR VENETO 2014-2020, and by the project SPIN 2021 “Requirement specification and static analysis of robotic software” - Ca’Foscari University. Any opinions, findings, conclusions, or recommendations expressed in this material are those of the authors and may not reflect those of the funding organizations.}%
\thanks{$^{1}$Alias Robotics,
        Venta de la Estrella 6, pab 130, Vitoria 01006, Spain
        {\tt\small victor@aliasrobotics.com}}%
\thanks{$^{2}$System Security Group, Universit\"at Klagenfurt, Universitätsstr. 65-67 9020 Klagenfurt, Austria
        {\tt\small v1mayoralv@edu.aau.at}}%
\thanks{$^{3}$Contextual Robotics Institute, UC San Diego,
{\tt\small rwhitema@ucsd.edu}}%
\thanks{$^{4}$Ca' Foscari University of Venice, Dorsoduro 3246, Venice 30123, Italy
{\tt\small gianluca.caiazza@unive.it}}%
\thanks{$^{5}$Secura Factors srls, via Torino 155, Venice 30170, Italy
{\tt\small info@securafactors.com}}%
\thanks{$^{6}$NeoFarm, Chemin des Quarante Arpents, 78860 Saint-Nom-la-Breteche, France
{\tt\small mikael.arguedas@gmail.com}}%
}

\maketitle
\begin{abstract}

ROS 2 is rapidly becoming a standard in the robotics industry. Built upon DDS as its default communication middleware and used in safety-critical scenarios, adding security to robots and ROS computational graphs is increasingly becoming a concern. The present work introduces SROS2, a series of developer tools and libraries that facilitate adding security to ROS 2 graphs. Focusing on a usability-centric approach in SROS2, we present a methodology for securing graphs systematically while following the DevSecOps model. We also demonstrate the use of our security tools by presenting an application case study that considers securing a graph using the popular Navigation2 and SLAM Toolbox stacks applied in a TurtleBot3 robot. We analyse the current capabilities of SROS2 and discuss the shortcomings, which provides insights for future contributions and extensions. Ultimately, we present SROS2 as usable security tools for ROS 2 and argue that without usability, security in robotics will be greatly impaired.

\end{abstract}


\section{Introduction}
\label{sec:introduction}

A robot is a network of networks\cite{mayoral2021hacking}. One that is comprised of sensors to perceive the world, actuators to produce a physical change, and computational resources to process it all and respond coherently, in time, and according to its application. Security is of paramount importance in this context, as any disruption of any of these robot networks can cause the complete robotic system to misbehave and compromise the safety of humans, as well as the environment \cite{kirschgens2018robot,zamalloa2017dissecting}. 

The Robot Operating System (ROS) \cite{quigley2009ros} is the \emph{de facto} framework for robot application development. Widely used to govern interactions across robot networks, at the time of writing, the original ROS article \cite{quigley2009ros} has been cited more than 9300 times, which shows its wide acceptance for research and academic purposes. ROS was born in this environment: its primary goal was to provide the software tools and libraries that users would need to employ to undertake novel robotics research and development. Adoption in industry has also been rapidly increasing over the last few years. According to the latest ROS community metrics \cite{rosmetrics} sampled every year in July, the number of ROS downloads has increased by over 50\%, with about 600 million downloads between July of 2020 and July of 2021. Moreover, based on the download percentages reported from \texttt{packages.ros.org}, we observe a significant increase in adopting ROS 2, which suggests that by 2023 there will be more users using ROS 2 than its predecessor\footnote{We also note that past studies estimated that by 2024, 55\% of the total commercial robots shipped that year will include at least one ROS package. For more details, refer to \emph{\url{https://www.businesswire.com/news/home/20190516005135/en/Rise-ROS-55-total-commercial-robots-shipped}}.}.

ROS was not designed initially with security in mind, but as it started being adopted and deployed into products or used in government programs, more attention was drawn to security issues. Some of the early work on securing ROS included \cite{lera2016ciberseguridad, ApplicationSecROS} or \cite{white2016sros}, both appearing in the second half of 2016. At the time of writing, none of these efforts remain actively maintained and the community focus on security efforts has switched to ROS 2. A recent study \cite{mayoral2022robot} that surveyed the security interests in the ROS community presented data indicating that 73\% of the survey participants considered that they had not invested enough to protect their robots from cyber-threats. The same number of participants indicated that their organizations were open to invest, however only 26\% acknowledged to actually have invested. This led the authors to conclude that there is a gap between the security expectations and the actual investment. We argue that this gap is a result not only of the immaturity of security in robotics or the know-how but also by the lack of usability of the available security tools. Being conscious that security in robotics is not a product, but a process that needs to be continuously assessed in a periodic manner \cite{mayoral2022robotteardown,zhu2021cybersecurity,mayoral2020alurity}, we advocate for a usable security approach in robotics as the best way to remain secure.

In this article we introduce SROS2, a series of developer tools, meant to be usable and that facilitate adding security capabilities to ROS 2 computational graphs. We present a security methodology consisting of six steps that allow securing ROS 2 graphs iteratively, with the aid of SROS2. Driven by an application use case, we discuss how SROS2 allows achieving security in complex graphs involving popular ROS 2 packages and analyze the security trade-offs and limitations of our current tooling. The key contributions of this work are:
\begin{itemize}
    \item Create SROS2, a set of usable tools for adding security to ROS 2 that: (1) help introspect the computational graph by extracting communication middleware-level information; (2) simplify the security operations creating Identity and Permissions Certificate Authorities (CA) that govern the security policies of a ROS 2 graph; (3) help organize all security artifacts in a consistent manner and within a directory tree that is generated within the current ROS 2 workspace overlay; (4) help create a new identity for each enclave, generating a keypair and signing its x.509 certificate using the appropriate CA; (5) create governance files to encrypt all DDS traffic by default; (6) support specifying enclave permissions in familiar ROS 2 terms which are then automatically converted into low-level DDS permissions; (7) support automatic discovery of required permissions from a running ROS 2 system; and (8) dissect communication middleware interactions, to extract key information for the security monitoring of the system.
    \item Propose a methodology for securing ROS 2 computational graphs that provides roboticists with a structured process to continuously assess their security.
    \item Expose insights into how to apply SROS2 to real ROS 2 computational graphs by presenting an application case study focused on analyzing the Navigation2 and SLAM Toolbox stacks in a TurtleBot3 robot.
\end{itemize}

The core components of SROS2 are disclosed under a commercially friendly open-source license and are available and maintained at \emph{\url{https://github.com/ros2/sros2}}.

\section{Related Work}
\label{sec:related_work}

Considering how ROS was originally intended as a fast prototyping robotic framework, security was not considered a priority feature in its first iteration. As ROS has evolved from the prototyping to the real-world industrial applications the entire stack came to be in dire need of cybersecurity attention \cite{mayoral2022robot} \cite{phdCaiazza}.  

A first partial analysis, with the goal of understanding what prevented ROS from being used industrially, was conducted by McClean \emph{et al.} \cite{McClean2013Preliminary}. By means of a 'honeypot' system, at DEFCON-20, they collected how malicious users would tackle a robot in the wild. Dieber \emph{et al.} \cite{dieber2019penetration} provided a complete and in-depth analysis of the security vulnerabilities and attack surfaces in ROS systems and how to exploit them, highlighting the gaps in the security of the framework. A considerable amount of research has been done as regards the publish-subscribe paradigm reviewing the performance and the techniques to secure it either via the communication channel, and ROS' internal mechanisms \cite{Goerke2021}. In the first case, via message authentication \cite{Toris2014Message}, within the later addition of using of encryption and security artifacts \cite{Lera2016Cybersecurity} \cite{francisco2018message} \cite{breiling2017secure}.
In the latter case, by enhancing the middleware behaviour with some extra, such as a run-time monitor to filter out and log all the requests and operations sent in the graph \cite{Huang2014ROSRV}, an Application Level Gateway - that wraps the existing API calls to enforce authentication and authorization - that exposes a permission token to be evaluated before executing the requested operation \cite{Doczi2016Increasing}; to the extent of forking the implementation, modifying the transport mode via IPSec \cite{Sundaresan2017SecureROS}, or via a security architecture intended with the addition of x.509 certificates and authorization server \cite{ApplicationSecROS}. Unfortunately, those approaches suffered to some extent with limitations and downsides, such as a lack of flexibility (e.g. Single Point of Failure (SPOF)) and usability, which were tackled in the Secure ROS (SROS) initiative \cite{white2018sros1}. With the objective of providing additions to the ROS API and ecosystem to support modern cryptography and security measures, the project introduced new security features to the core of ROS’ codebase and, more importantly, a set of tooling to ease the burden on the developers of correctly implementing security.

With the second iteration of the framework in ROS 2, thanks to the adoption of DDS as the communication middleware\footnote{https://design.ros2.org/articles/ros\_on\_dds.html}, we observed how the inherited security measures and methodologies in the system have evolved the framework. However, we can no longer overlook how its complexity still remains prone to human error in processes such as the access control artifacts distribution \cite{white2018procedurally}, or even to overlooking exposed attack surfaces \cite{white2019network}. Moreover, keeping track of all the new pieces to the ROS puzzle became even more demanding and lengthy procedurally, requiring continuous attention and systematic security analysis--which left usability challenging \cite{phdWhite}. Our work addresses this challenge with a security toolset (SROS2) and a security methodology for robotics.

\label{subsec:todo}

\section{Approach}
\label{sec:approach}

\begin{figure}[!h]
    \centering
    \includegraphics[width=0.4\textwidth]{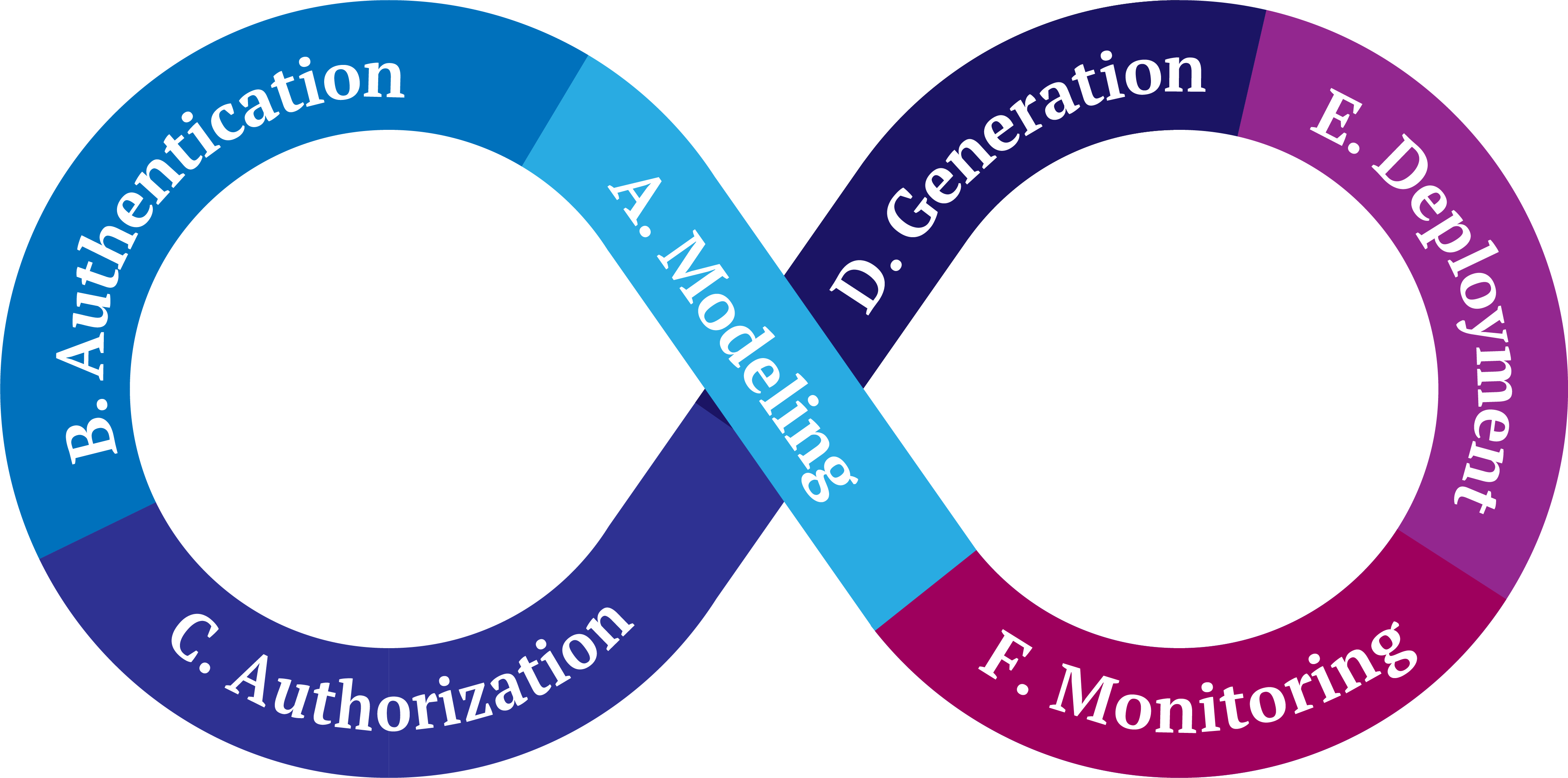}
    \caption{\textbf{Our methodology for securing ROS 2 computational graphs}. SROS2 provides tools and libraries to facilitate a secure DevOps model in robotics (DevSecOps \cite{mayoral2020devsecops}).}
    \label{fig:methodology}
\end{figure}

We propose the following methodology (Fig. \ref{fig:methodology}) to secure ROS 2 computational graphs:
(A) introspect the graph and model its security landscape to determine the necessary security policies and enclaves; implement such policies by (B) defining the authentication and (C) authorization configurations; (D) generate all the required security artifacts; (E) deploy them appropriately across robotic systems; and (F) continuously monitor the network, reverting to (A) modeling when appropriate.

\subsection{Modeling}
\label{approach:modeling}

Modeling refers to the use of abstractions to aid in a thought process. In security, threat modeling aids in thinking about risks and determines the threat landscape. The output of this effort is often called the threat model. Commonly, a threat model enumerates the potential attackers, their capabilities, resources and their intended targets. In the context of robot cybersecurity, a threat model identifies security threats that apply to the robot and/or its components\footnote{both software and hardware, including computational graph resources.} while providing means to address or mitigate them for a particular use case. A threat model also provides inputs that are used to then determine a set of policy rules (or principles) that direct how ROS 2 should provide security services to protect sensitive and critical graph resources. When put together these policy rules are called the \emph{security policy}.

SROS2 aims to provide tools to introspect and model the security of ROS 2 computational graphs into the desired security policies. Introspection of the graph can be performed in two ways:

\begin{enumerate}
    \item By leveraging the ROS 2 API and the framework for ROS 2 command line tools (\texttt{ros2cli}), we can pull ROS \emph{Nodes}, \emph{Topics}, \emph{Services} or \emph{Actions} information (among others) from the ROS 2 graph and display these in the CLI, see Listing \ref{lst:modeling}. This allows us to get a grasp of the computational graph from a ROS 2 perspective. Other tools such as \texttt{RViz} \cite{kam2015rviz} or \texttt{rqt} help get a visual depiction of the graph and its abstractions.
    \item Monitoring network interactions at the ROS communication middleware-level can be extremely helpful to model security but incredibly cumbersome from a usability perspective unless the right tooling is provided. ROS 2 uses OMG's Data Distribution Service (DDS) \cite{dds14} as its default communication middleware, which is a complex specification. To facilitate introspection of DDS, SROS2 leverages \emph{scapy} \cite{rohith2018scapy}, a powerful interactive packet manipulation library that allows us to forge or decode network packets. Particularly, we contributed an open source \emph{scapy} dissector\footnote{see \url{https://github.com/secdev/scapy/pull/3403}} that allows us to dissect the wire-level communication protocol that is used by the default ROS 2 communication middleware: the Real-Time Publish Subscribe protocol (RTPS) \cite{rtps25}. Using this, SROS2 provides tooling that allows monitoring network interactions, capturing DDS databus information directly and displaying these for the security analyst's consumption.
\end{enumerate}

\lstset{language=bash}
\lstset{label={lst:modeling}}
\lstset{basicstyle=\ttfamily\footnotesize,
    commentstyle=\color{MidnightBlue}}
\lstset{caption={SROS2 extends ROS 2 APIs to facilitate computational graph introspection at the networking level for modeling purposes.} 
}
\lstset{escapeinside={<@}{@>}}
\begin{lstlisting}
# ROS 2 CLI API allows direct introspection
ros2 topic list
<@\textcolor{gray}{/cmd\_vel}@> <@\tikz[remember picture] \node [] (p1) {}; @>
<@\textcolor{gray}{/robot\_state\_publisher}@> <@\tikz[remember picture] \node [] (p1) {}; @>
...
ros2 node list
<@\textcolor{gray}{/turtlebot3\_diff\_drive}@> <@\tikz[remember picture] \node [] (p1) {}; @>
...
# SROS2 extensions allow introspecting DDS
ros2 security introspection
<@\textcolor{gray}{DDS endpoint detected (hostId=16974402, appId=2886795267, instanceId=10045242)}@> <@\tikz[remember picture] \node [] (p1) {}; @>
    <@\textcolor{gray}{- version: 2.4}@> <@\tikz[remember picture] \node [] (p1) {}; @>
    <@\textcolor{gray}{- vendorId: ADLINK - Cyclone DDS}@> <@\tikz[remember picture] \node [] (p1) {}; @>
    <@\textcolor{gray}{- IP: 192.168.1.34}@> <@\tikz[remember picture] \node [] (p1) {}; @>
    <@\textcolor{gray}{- transport: UDP}@> <@\tikz[remember picture] \node [] (p1) {}; @>
<@\textcolor{gray}{DDS endpoint detected (...)}@>
\end{lstlisting}

For complete threat modeling, we refer the reader to \cite{ros2securitytm} which discusses details around security modeling ROS 2 computational graphs.

\subsection{Authentication}
\label{approach:authentication}
Authentication provides proof of a claimed identity ($\neq$ identification,  determination of an unknown entity). ROS 2 offloads authentication to its underlying communication middleware, DDS. By default, DDS allows any arbitrary \emph{DomainParticipant} to join any \emph{Domain} without authentication. DDS however provides the means to verify the identity of the application and/or the user that invokes operations on the DDS databus through its \emph{DDS Security} extensions \cite{ddssecurity11}.  With these, for protected DDS \emph{Domains}, a \emph{DomainParticipant} that enables authentication will only communicate with other \emph{DomainParticipants} that also have authentication enabled. 

To favour usability and reduce human errors, all implementation details of authentication in ROS 2 through DDS are abstracted away by our SROS2 tools. The appropriate artifacts for enabling authentication capabilities are produced in the \emph{Generation} step (\ref{approach:generation}) of our methodology, and default to the security mechanisms specified by OMG's DDS Security \cite{ddssecurity11}. In particular, each \emph{DomainParticipant} uses a Public Key Infrastructure (PKI) with a common shared Certificate Authority (CA): \texttt{Identity CA}. All participants interoperating securely must be pre-configured with \texttt{Identity CA} and have a signed certificate from it. Participants are expected to use mutual authentication through a challenge-response mechanism supported by either the Rivest Shamir Adleman (RSA) \cite{rivest1983cryptographic} or the Elliptic Curve Digital Signature Algorithm (ECDSA) \cite{johnson2001elliptic} asymmetric encryption algorithms. Shared secrets are established using using the  Diffie-Hellman (DH) \cite{diffie1976new} or Elliptic Curve DH (ECDH) (Ephemeral Mode) \cite{merkle1978secure} key agreement protocols.

Listing \ref{lst:authentication} shows an example of how SROS2 tools abstract the complexity of DDS authentication away from ROS developers. The \texttt{governance.xml} policy document is auto-generated by SROS2 and captures domain-wide security settings that include authentication aspects. Additional details about the underlying authentication process and the security artifacts are available in \cite{ddssecurity11}, \cite{ros2securityintegration} and \cite{ros2securityenclaves}. 

\lstset{language=xml}
\lstset{label={lst:authentication}}
\lstset{basicstyle=\ttfamily\footnotesize,
    commentstyle=\color{blue},
    stringstyle=\ttfamily\color{red!50!brown},}
\lstset{caption={An extract from \texttt{governance.xml} policy document generated by SROS2 illustrating domain-wide security settings such as how to handle unauthenticated participants, whether to encrypt discovery or the default rules for access to topics.} 
}
\lstset{escapeinside={<@}{@>}}
\begin{lstlisting}
...
<<@\textcolor{Plum}{allow\_unauthenticated\_participants}@>><@\textcolor{ForestGreen}{false}@></<@\textcolor{Plum}{allow\_unauthenticated\_participants}@>>
<<@\textcolor{Plum}{enable\_join\_access\_control}@>><@\textcolor{ForestGreen}{true}@></<@\textcolor{Plum}{enable\_join\_access\_control}@>>
<<@\textcolor{Plum}{discovery\_protection\_kind}@>><@\textcolor{ForestGreen}{ENCRYPT}@></<@\textcolor{Plum}{discovery\_protection\_kind}@>>
<<@\textcolor{Plum}{liveliness\_protection\_kind}@>><@\textcolor{ForestGreen}{ENCRYPT}@></<@\textcolor{Plum}{liveliness\_protection\_kind}@>>
<<@\textcolor{Plum}{rtps\_protection\_kind}@>><@\textcolor{ForestGreen}{SIGN}@></<@\textcolor{Plum}{rtps\_protection\_kind}@>>
<<@\textcolor{Plum}{topic\_access\_rules}@>>
    <<@\textcolor{Mulberry}{topic\_rule}@>>
        <<@\textcolor{Periwinkle}{topic\_expression}@>><@\textcolor{ForestGreen}{*}@></<@\textcolor{Periwinkle}{topic\_expression}@>>
        <<@\textcolor{Periwinkle}{enable\_discovery\_protection}@>><@\textcolor{ForestGreen}{true}@></<@\textcolor{Periwinkle}{enable\_discovery\_protection}@>>
        <<@\textcolor{Periwinkle}{enable\_liveliness\_protection}@>><@\textcolor{ForestGreen}{true}@></<@\textcolor{Periwinkle}{enable\_liveliness\_protection}@>>
        <<@\textcolor{Periwinkle}{enable\_read\_access\_control}@>><@\textcolor{ForestGreen}{true}@></<@\textcolor{Periwinkle}{enable\_read\_access\_control}@>>
        <<@\textcolor{Periwinkle}{enable\_write\_access\_control}@>><@\textcolor{ForestGreen}{true}@></<@\textcolor{Periwinkle}{enable\_write\_access\_control}@>>
        <<@\textcolor{Periwinkle}{metadata\_protection\_kind}@>><@\textcolor{ForestGreen}{ENCRYPT}@></<@\textcolor{Periwinkle}{metadata\_protection\_kind}@>>
        <<@\textcolor{Periwinkle}{data\_protection\_kind}@>><@\textcolor{ForestGreen}{ENCRYPT}@></<@\textcolor{Periwinkle}{data\_protection\_kind}@>>
    </<@\textcolor{Mulberry}{topic\_rule}@>>
</<@\textcolor{Plum}{topic\_access\_rules}@>>
...
\end{lstlisting}

\subsection{Authorization}
\label{approach:authorization}
Authorization helps define and verify the policies that are assigned to a certain identity. Access control instead --also called permissions or privileges-- are the methods used to enforce such policies. While access control is handled by the DDS implementation, authorization policies need to be defined by the developer. SROS2 helps map these policies from the ROS 2 computational graph to the underlying DDS databus abstractions through two resources: the \texttt{Permissions CA} and a \texttt{permissions.xml} policy document. Listing \ref{lst:authorization} shows an extract from one of the policy documents that defines the authorization profile for a particular ROS 2 Node. Details about how access control is implemented by the underlying communication middleware are discussed in \cite{ddssecurity11} and \cite{ros2securityaccesscontrol}.

\lstset{language=xml}
\lstset{label={lst:authorization}}
\lstset{basicstyle=\ttfamily\footnotesize,
    commentstyle=\color{blue},
    stringstyle=\ttfamily\color{red!50!brown},}
\lstset{caption={SROS2 provides means to define authentication policies through XML files.} 
}
\lstset{escapeinside={<@}{@>}}
\begin{lstlisting}
<profile node="turtlebot3_diff_drive" ns="/">
    <xi:include href="common/node.xml"
      xpointer="xpointer(/profile/*)"/>
    <topics subscribe="ALLOW">
      <topic><@\textcolor{ForestGreen}{/cmd\_vel}@></topic>
    </topics>
    <topics publish="ALLOW">
      <topic><@\textcolor{ForestGreen}{odom}@></topic>
      <topic><@\textcolor{ForestGreen}{tf}@></topic>
    </topics>
</profile>
<profile node="turtlebot3_imu" ns="/">
...
</profile>
\end{lstlisting}

\subsection{Generation}
\label{approach:generation}

\emph{Modeling} (\ref{approach:modeling}), \emph{Authentication} (\ref{approach:authentication}) and \emph{Authorization} (\ref{approach:authorization}) steps of our methodology (Fig. \ref{fig:methodology}) help define one or multiple security policies. To help implement such policies, SROS2 provides means to automate the generation of the corresponding security artifacts and simplify the translation to the underlying DDS implementation. To do so, SROS2 maps a security policy to an \emph{enclave}: a set of ROS 2 computational graph resources that operate in the same security domain, use particular \texttt{Identity CA} and \texttt{Permissions CA} authorities, and share the protection of a single, common, continuous security perimeter. 

All secure interactions in ROS 2 computational graphs must use an enclave that contains the runtime security artifacts unique to that enclave, yet each Node may not necessarily have a unique enclave. Multiple enclaves can be encapsulated in a single security policy to accurately model the information flow control. Users can then tune the fidelity of such models by controlling at what scope enclaves are applied at deployment. For example, one unique enclave per robot, or per swarm, or per network, etc.

Listing \ref{lst:generation} shows how SROS2 tools help generate all artifacts to implement a new security policy, inferred directly from the running ROS 2 graph. For a more complex policy that involves multiple enclaves, we refer the reader to \url{https://github.com/ros-swg/turtlebot3\_demo/blob/2719e0f/policies/tb3\_gazebo\_policy.xml}.

\lstset{language=bash}
\lstset{label={lst:generation}}
\lstset{basicstyle=\ttfamily\footnotesize,
    commentstyle=\color{MidnightBlue}}
\lstset{caption={SROS2 provides tools to implement security policies and in ROS 2 computational graphs, generating all security artifacts necessary.} 
}
\lstset{escapeinside={<@}{@>}}
\begin{lstlisting}
# Generate a new keystore with Identify and Permission CA keys, associated certificates and a authentication structure through governance.xml file
ros2 security create_keystore new_keystore

# Inspect current ROS graph and produce a security policy
ros2 security generate_policy new_keystore/my_policy.xml

# Generate all security artifacts necessary to enforce the policy, this includes enclaves and the access control permission files
ros2 security generate_artifacts \
  -k new_keystore \
  -p new_keystore/my_policy.xml 
\end{lstlisting}

\subsection{Deployment}
\label{approach:deployment}

Deployment is a relevant phase in the methodology of Fig. \ref{fig:methodology} and must be also exercised securely. We consider three types of deployments of both artifacts and secure information:

\begin{enumerate}
    \item \underline{Distribution of policy artifacts}: the resulting artifacts from the \emph{Generation phase} (\ref{approach:generation}) must be securely deployed into the targeted robots and related systems. At the time of writing SROS2 does not provide any particular special utilities to deploy security artifacts. We however direct readers to the ongoing efforts to launch ROS 2 graphs remotely and in multi-machine environments \cite{multimachinedeployros2} for inspiration.
    \item \underline{Message authentication}: verification of the Message Authentication Codes (MAC) is performed using Advanced Encryption Standard (AES) with Galois MAC (AES-GMAC). DDS security extensions abstract this away from the ROS developer.
    \item \underline{Encryption of secure DDS interactions}: authenticated symmetric cryptography governs all DDS interactions within a security policy using also AES in Galois Counter Mode (AES-GCM). Similar to message authentication, DDS abstracts this away from the developer and is enabled automatically provided that the security policy is configured appropriately.
\end{enumerate}

\subsection{Monitoring and mitigation}
\label{approach:monitoring}

The last phase in our methodology leads to a never ending loop of continuous \emph{Monitoring, mitigation} (\ref{approach:monitoring}) and \emph{Modeling} (\ref{approach:modeling}). This way, security in ROS 2 computational graphs becomes a moving target, a process--one that demands continuous assessments as changes occur in the robots, the network, or as new security flaws are discovered affecting the running systems.

SROS2 provides tools for monitoring running ROS 2 graphs and detecting possible flaws. Listing \ref{lst:monitoring} shows an example:

\begin{figure*}[!htb]
    \centering
    \includegraphics[
        width=\linewidth,
        trim= 0 300 1500 0,
        clip
        ]
        {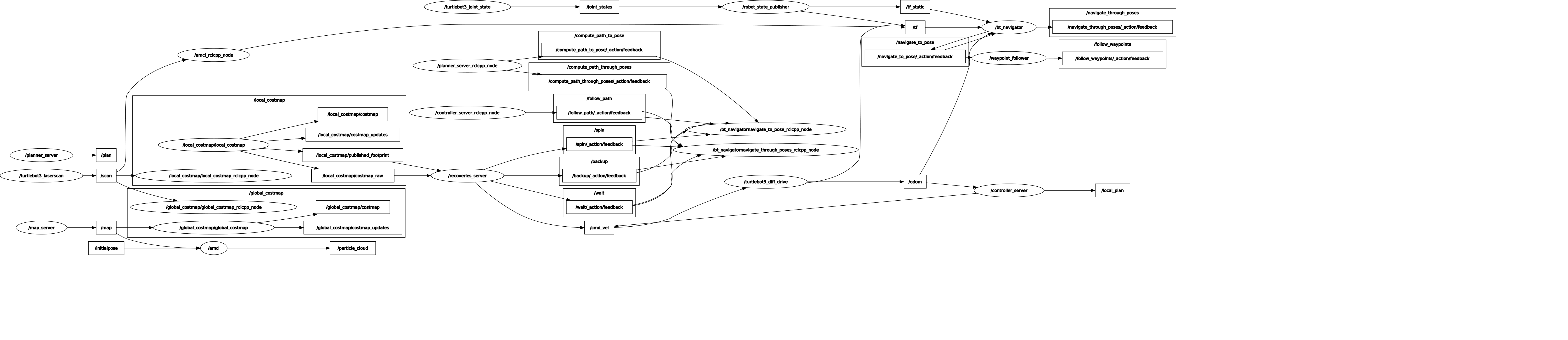}
    \caption{\textbf{A portion of the simulated TurtleBot3 secured using SROS2}. The figure depicts a subset of the computational graph of the robot including sensor and control topics as well some relevant portions of the \texttt{navigation2} software stack.}
    \label{fig:navigationgraph}
\end{figure*}


\lstset{language=bash}
\lstset{label={lst:monitoring}}
\lstset{basicstyle=\ttfamily\footnotesize,
    commentstyle=\color{MidnightBlue}}
\lstset{caption={SROS2 provides tools to dissect DDS interactions, extract key information and map it to outstanding security flaws affecting DDS.} 
}
\lstset{escapeinside={<@}{@>}}
\begin{lstlisting}
# monitor the network segment for vulnerabilities affecting DDS participants
ros2 security monitor
<@\textcolor{gray}{sniffing the DDS network...}@>
<@\textcolor{RedViolet}{Vulnerable DDS endpoint found (hostId=16974402, appId=2886795267, instanceId=10045242)}@>
    <@\textcolor{RedViolet}{- vendorId: Real-Time Innovations, Inc. - Connext DDS}@>
    <@\textcolor{RedViolet}{- version: 6.0.1.25}@>
    <@\textcolor{RedViolet}{- CVE IDs:}@>
        <@\textcolor{RedViolet}{* CVE-2021-38487}@>
        <@\textcolor{RedViolet}{* CVE-2021-38435}@>
\end{lstlisting}
\section{Application and analysis}
\label{sec:application}

To apply our methodology, as defined in Fig. \ref{fig:methodology}, we demonstrate the application of SROS2 using two of the most commonly used frameworks in ROS 2, the \texttt{navigation2} \cite{macenski2020marathon2} and \texttt{slam\_toolbox} \cite{macenski2021slamtoolbox} stacks. Particularly, the Navigation2 project\footnote{\url{https://github.com/ros-planning/navigation2}} provides a software stack including path planning algorithms and behavioral navigation servers that can be seamlessly integrated with existing sensor perception pipelines, localization and mapping services, and drivetrain velocity controllers to support various mobile robotic applications. While Navigation2 remains mostly agnostic of robotic platforms, we selected the widely accessible and community supported TurtleBot3\footnote{\url{https://www.robotis.us/turtlebot-3}} as our target robot for analysis--consisting of a differential drive, circular base footprint, and ground level 2D scanning LIDAR. Our application case study is depicted in Fig. \ref{fig:navigationgraph}.

To start, we begin with the modeling (\ref{approach:modeling}), authentication (\ref{approach:authentication}) and authorization (\ref{approach:authorization}) phases in order to bootstrap an initial security policy that captures the minimal spanning set of security measures required for the nominal function of the application across the distributed computation graph. We can either first bring up the ROS 2 application under a controlled network environment with security mode disabled, or provision an initial keystore enclave with temporary key-material and only access control governance disabled; the first option includes minimal setup while being more transparent to debug, while the later is advantageous in modeling policies directly from field deployments across untrusted networks.

With the ROS 2 application running, SROS2 can capture the topology of the computation graph to populate the permission profiles within our initial policy, registering each active ROS 2 node and respective topic publication and subscription. A limitation in SROS2’s current snapshot approach however is in accurately modeling more ephemeral resource access events, such as service clients or action requests. While ROS 2’s internal graph API (that SROS2 uses to sample topology measurements) provides a middleware agnostic interface, the observation window is only instantaneous and can easily miss asynchronous resource access events.

Given the graph API limitations, it’s often necessary to iteratively test the generated policy by using it to update the signed permission and governance files and relaunch the application with access control enabled. For moderate to advanced applications such as those relying on Navigation2, permission access denied errors may inevitably be encountered. With ROS 2 however, such events can be logged and aggregated into policy refinement, specifying the node and resource namespaces denied.

After iterative policy refinement, once the tested application is fully functional with enforced access control, the policy can then be further optimized. Such policy optimizations include sorting common permission patterns into smaller sub-profiles, being more manageable to audit and modularly reusable across repeating permission sets in a global policy. We demonstrate it in \emph{\url{https://github.com/ros-swg/turtlebot3_demo/tree/master/policies}}. This auditing process also provides an opportunity to assess the granularity of the policy as well, from both permission Access Control (AC) and Information Flow Control (IFC) perspectives.

While the minimal spanning set of AC permissions may be optimally secure in terms of the Principle of Least Privilege \cite{schneider2003least}, it may not be optimally usable for a target application domain. Though most computation graphs in ROS 2 are largely static at runtime, cases where resource namespaces change over the application’s lifecycle do exist. For example, multi-robot systems may fluctuate as agents enter or exit networks for missions or maintenance. Additionally, node namespaces sometimes include sequence numbers to ensure namespace uniqueness. To accommodate such scenarios, permissions could be modified to include wildcards as necessary. While static permissions are straightforward to interpret and less likely to inadvertently introduce policy flaws, wildcarding select permissions provides a usable compromise when required.

When auditing from an IFC perspective, optimizing the policy into assorted enclaves becomes a key consideration. As all ROS 2 nodes composed into a shared process share a common DDS context, they subsequently share the same security enclave or set of permission profiles. This of course is inherently coupled with how the application is architected and to be deployed across a distributed system. As such, security requirements for IFC may then instead dictate aspects of the application's designs. The degree of granularity of IFC sought then dictates the allotment of enclaves used to contain sub-profiles for the application’s policy.

In the case of Navigation2 and its large degree of coupling and composition of nodes, the planning stack derived from a single source tree is perhaps best relegated to its own enclave, while still being readily separable from any other enclave dedicated to perception or control nodes. Admittedly, such auditing procedures in determining the allotment of enclaves remains rather ambiguous for users, and so presents another area of ergonomics for SROS2 to help automate or advise through formal analysis.

The source code of our application case study is available at \emph{\url{https://github.com/ros-swg/turtlebot3_demo}}. The resulting security policies of applying our methodology (Fig. \ref{fig:methodology}) are also available in the same repository and show various profiles that result from a systematic assessment.

\section{Conclusion}
\label{sec:conclusion}

In this work we present SROS2, a series of developer tools focused on usable security that allow adding security capabilities to ROS 2 computational graphs. We introduce a methodology around these tools consisting of 6 basic steps depicted in Fig. \ref{fig:methodology} and aligning to the common DevSecOps flows: (A) introspect the computational graph and model its security to determine the necessary security policies and enclaves; (B) define authentication and (C) authorization configurations; (D) generate all the required security artifacts for implementing such policies; (E) deploy them appropriately across robotic systems; and (F) continuously monitor the network, reverting to (A) modeling when needed. SROS2 facilitates each one of these steps by integrating itself tightly into the usual ROS 2 development flows.

We present an application case study discussing how to propose a secure architecture for the TurtleBot3 robot using the \texttt{navigation2} and \texttt{slam\_toolbox} stacks. This is of special interest since it aligns to the software architecture that many industrial and professional robots are using today, given the popularity of these packages.

We introduce security as a process in robotics and correspondingly, our work aims to pave the way for enabling security processes, particularly in ROS 2. Alongside the never-ending reality of security, we acknowledge that SROS2 has various limitations that deserve further attention and improvements. Some of these include the lack of granularity of security configurations in the current abstractions, which makes it difficult to configure encryption and authentication options separately. Others refer to the lifecycle management of security artifacts, including updating certificates and keys, wherein secure deployment plays a key role. We are particularly keen on improving SROS2 mechanisms in the future to ensure secure lifecycles while minimizing the downtime impact in ROS 2 graphs. 
Promising directions for future work also include the development of more advanced monitoring and introspection capabilities, the extension of SROS2 to other communication middlewares (beyond DDS) and finally, the continuous improvement of the usability of the tools. For this, we believe that the use of Graphical User Interfaces (GUIs) represents an interesting opportunity to further facilitate SROS2 usability to non-roboticists.

Our work aims to inspire groups in robotics to add security to their robotic computational graphs. We look forward to security in robotics becoming more usable and accessible, minimizing the threat landscape that lies before us now, and closing the window of opportunity for bad actors.

\newpage


\bibliographystyle{IEEEtran}
\bibliography{references}

\begin{thebibliography}{10}
\providecommand{\url}[1]{#1}
\csname url@samestyle\endcsname
\providecommand{\newblock}{\relax}
\providecommand{\bibinfo}[2]{#2}
\providecommand{\BIBentrySTDinterwordspacing}{\spaceskip=0pt\relax}
\providecommand{\BIBentryALTinterwordstretchfactor}{4}
\providecommand{\BIBentryALTinterwordspacing}{\spaceskip=\fontdimen2\font plus
\BIBentryALTinterwordstretchfactor\fontdimen3\font minus
  \fontdimen4\font\relax}
\providecommand{\BIBforeignlanguage}[2]{{%
\expandafter\ifx\csname l@#1\endcsname\relax
\typeout{** WARNING: IEEEtran.bst: No hyphenation pattern has been}%
\typeout{** loaded for the language `#1'. Using the pattern for}%
\typeout{** the default language instead.}%
\else
\language=\csname l@#1\endcsname
\fi
#2}}
\providecommand{\BIBdecl}{\relax}
\BIBdecl

\bibitem{mayoral2021hacking}
V.~Mayoral-Vilches, A.~Glera-Pic{\'o}n, U.~Ay{\'u}car-Carbajo, S.~Rass,
  M.~Pinzger, F.~Maggi, and E.~Gil-Uriarte, ``Hacking planned obsolescense in
  robotics, towards security-oriented robot teardown,'' \emph{Electronic
  Communications of the EASST}, vol.~80, 2021.

\bibitem{kirschgens2018robot}
L.~A. Kirschgens, I.~Z. Ugarte, E.~G. Uriarte, A.~M. Rosas, and V.~M. Vilches,
  ``Robot hazards: from safety to security,'' \emph{arXiv preprint
  arXiv:1806.06681}, 2018.

\bibitem{zamalloa2017dissecting}
I.~Zamalloa, R.~Kojcev, A.~Hern{\'a}ndez, I.~Muguruza, L.~Usategui, A.~Bilbao,
  and V.~Mayoral, ``Dissecting robotics-historical overview and future
  perspectives,'' \emph{arXiv preprint arXiv:1704.08617}, 2017.

\bibitem{quigley2009ros}
M.~Quigley, K.~Conley, B.~Gerkey, J.~Faust, T.~Foote, J.~Leibs, R.~Wheeler, and
  A.~Y. Ng, ``Ros: an open-source robot operating system,'' in \emph{ICRA
  workshop on open source software}, vol.~3, no. 3.2.\hskip 1em plus 0.5em
  minus 0.4em\relax Kobe, Japan, 2009, p.~5.

\bibitem{rosmetrics}
\BIBentryALTinterwordspacing
R.~community, ``Ros community metrics,'' 2021. [Online]. Available:
  \url{http://wiki.ros.org/Metrics}
\BIBentrySTDinterwordspacing

\bibitem{lera2016ciberseguridad}
F.~J.~R. Lera, V.~Matell{\'a}n, J.~Balsa, and F.~Casado, ``Ciberseguridad en
  robots aut{\'o}nomos: An{\'a}lisis y evaluaci{\'o}n multiplataforma del
  bastionado ros,'' \emph{Actas Jornadas Sarteco}, pp. 571--578, 2016.

\bibitem{ApplicationSecROS}
B.~Dieber, S.~Kacianka, S.~Rass, and P.~Schartner, ``Application-level security
  for ros-based applications,'' in \emph{2016 IEEE/RSJ International Conference
  on Intelligent Robots and Systems (IROS)}, Oct 2016, pp. 4477--4482.

\bibitem{white2016sros}
R.~White, M.~Quigley, and H.~Christensen, ``{SROS}: Securing {ROS} over the
  wire, in the graph, and through the kernel,'' in \emph{Humanoids Workshop:
  Towards Humanoid Robots {OS}}.\hskip 1em plus 0.5em minus 0.4em\relax Cancun,
  Mexico, 2016.

\bibitem{mayoral2022robot}
V.~Mayoral-Vilches, ``Robot cybersecurity, a review,'' \emph{International
  Journal of Cyber Forensics and Advanced Threat Investigations}, 2022.

\bibitem{mayoral2022robotteardown}
V.~Mayoral-Vilches, A.~Glera-Pic{\'o}n, U.~Ayucar-Carbajo, S.~Rass, M.~Pinzger,
  F.~Maggi, and E.~Gil-Uriarte, ``Robot teardown, stripping industrial robots
  for good,'' \emph{International Journal of Cyber Forensics and Advanced
  Threat Investigations}, 2022.

\bibitem{zhu2021cybersecurity}
Q.~Zhu, S.~Rass, B.~Dieber, and V.~M. Vilches, ``Cybersecurity in robotics:
  Challenges, quantitative modeling, and practice,'' \emph{arXiv preprint
  arXiv:2103.05789}, 2021.

\bibitem{mayoral2020alurity}
V.~Mayoral-Vilches, I.~Abad-Fern{\'a}ndez, M.~Pinzger, S.~Rass, B.~Dieber,
  A.~Cunha, F.~J. Rodr{\'\i}guez-Lera, G.~Lacava, A.~Marotta, F.~Martinelli
  \emph{et~al.}, ``alurity, a toolbox for robot cybersecurity,'' \emph{arXiv
  preprint arXiv:2010.07759}, 2020.

\bibitem{phdCaiazza}
G.~Caiazza, ``Application-level security for robotic networks,'' Ph.D.
  dissertation, Ca' Foscari University of Venice, Italy, 2021.

\bibitem{McClean2013Preliminary}
\BIBentryALTinterwordspacing
J.~McClean, C.~Stull, C.~Farrar, and D.~Mascare{\~{n}}as, ``{A preliminary
  cyber-physical security assessment of the Robot Operating System (ROS)},''
  vol. 8741, p. 874110, 2013. [Online]. Available:
  \url{http://proceedings.spiedigitallibrary.org/proceeding.aspx?doi=10.1117/12.2016189}
\BIBentrySTDinterwordspacing

\bibitem{dieber2019penetration}
B.~Dieber, R.~White, S.~Taurer, B.~Breiling, G.~Caiazza, A.~Cortesi, and
  H.~Christensen, ``Penetration testing {ROS},'' in \emph{Robot Operating
  System ({ROS}): The Complete Reference (Volume 4)}.\hskip 1em plus 0.5em
  minus 0.4em\relax Springer International Publishing, 2019.

\bibitem{Goerke2021}
N.~Goerke, D.~Timmermann, and I.~Baumgart, ``Who controls your robot? an
  evaluation of ros security mechanisms,'' 02 2021, pp. 60--66.

\bibitem{Toris2014Message}
R.~Toris, C.~Shue, and S.~Chernova, ``Message authentication codes for secure
  remote non-native client connections to ros enabled robots,'' in \emph{2014
  IEEE International Conference on Technologies for Practical Robot
  Applications (TePRA)}, April 2014, pp. 1--6.

\bibitem{Lera2016Cybersecurity}
F.~J.~R. Lera, J.~Balsa, F.~Casado, C.~Fern{\'a}ndez, F.~M. Rico, and
  V.~Matell{\'a}n, ``Cybersecurity in autonomous systems: Evaluating the
  performance of hardening {ROS},'' \emph{M{\'a}laga, Spain-June 2016}, p.~47,
  2016.

\bibitem{francisco2018message}
\BIBentryALTinterwordspacing
F.~J. Rodr{\i}guez-Lera, V.~Matell{\'a}n-Olivera, J.~Balsa-Comer{\'o}n,
  {\'A}.~M. Guerrero-Higueras, and C.~Fern{\'a}ndez-Llamas, ``Message
  encryption in robot operating system: Collateral effects of hardening mobile
  robots,'' \emph{Frontiers in ICT}, vol.~5, p.~2, 2018. [Online]. Available:
  \url{https://www.frontiersin.org/article/10.3389/fict.2018.00002}
\BIBentrySTDinterwordspacing

\bibitem{breiling2017secure}
B.~Breiling, B.~Dieber, and P.~Schartner, ``Secure communication for the robot
  operating system,'' in \emph{2017 Annual IEEE International Systems
  Conference (SysCon)}, April 2017, pp. 1--6.

\bibitem{Huang2014ROSRV}
J.~Huang, C.~Erdogan, Y.~Zhang, B.~Moore, Q.~Luo, A.~Sundaresan, and G.~Rosu,
  ``Rosrv: Runtime verification for robots,'' in \emph{Proceedings of the 14th
  International Conference on Runtime Verification}, ser. LNCS, vol.
  8734.\hskip 1em plus 0.5em minus 0.4em\relax Springer International
  Publishing, September 2014, pp. 247--254.

\bibitem{Doczi2016Increasing}
R.~Dóczi, F.~Kis, B.~Sütő, V.~Póser, G.~Kronreif, E.~Jósvai, and
  M.~Kozlovszky, ``Increasing ros 1.x communication security for medical
  surgery robot,'' in \emph{2016 IEEE International Conference on Systems, Man,
  and Cybernetics (SMC)}, Oct 2016, pp. 4444--4449.

\bibitem{Sundaresan2017SecureROS}
\BIBentryALTinterwordspacing
A.~Sundaresan, L.~Gerard, and M.~Kim, ``Secure {ROS} 0.9.2 documentation,''
  July 2017. [Online]. Available: \url{https://sri-csl.github.io/secure_ros}
\BIBentrySTDinterwordspacing

\bibitem{white2018sros1}
R.~White, G.~Caiazza, A.~Cortesi, and H.~Christensen, ``{SROS1}: Using and
  developing secure ros1 systems,'' in \emph{Robot Operating System ({ROS}):
  The Complete Reference (Volume 3)}.\hskip 1em plus 0.5em minus 0.4em\relax
  Springer International Publishing, 2018.

\bibitem{white2018procedurally}
R.~{White}, H.~I. {Christensen}, G.~{Caiazza}, and A.~{Cortesi}, ``Procedurally
  provisioned access control for robotic systems,'' in \emph{2018 IEEE/RSJ
  International Conference on Intelligent Robots and Systems (IROS)}, 2018, pp.
  1--9.

\bibitem{white2019network}
R.~{White}, G.~{Caiazza}, C.~{Jiang}, X.~{Ou}, Z.~{Yang}, A.~{Cortesi}, and
  H.~{Christensen}, ``Network reconnaissance and vulnerability excavation of
  secure dds systems,'' in \emph{2019 IEEE European Symposium on Security and
  Privacy Workshops (EuroS PW)}, 2019, pp. 57--66.

\bibitem{phdWhite}
R.~White, ``Usable security and verification for distributed robotic systems,''
  Ph.D. dissertation, University of California San Diego, 2021.

\bibitem{mayoral2020devsecops}
V.~Mayoral-Vilches, N.~Garc{\'\i}a-Maestro, M.~Towers, and E.~Gil-Uriarte,
  ``Devsecops in robotics,'' \emph{arXiv preprint arXiv:2003.10402}, 2020.

\bibitem{kam2015rviz}
H.~R. Kam, S.-H. Lee, T.~Park, and C.-H. Kim, ``Rviz: a toolkit for real domain
  data visualization,'' \emph{Telecommunication Systems}, vol.~60, no.~2, pp.
  337--345, 2015.

\bibitem{dds14}
\BIBentryALTinterwordspacing
O.~M.~G. (OMG), ``Omg data distribution service (dds), version 1.4,'' 2015.
  [Online]. Available: \url{https://www.omg.org/spec/DDS/}
\BIBentrySTDinterwordspacing

\bibitem{rohith2018scapy}
R.~Rohith, M.~Moharir, G.~Shobha \emph{et~al.}, ``Scapy-a powerful interactive
  packet manipulation program,'' in \emph{2018 international conference on
  networking, embedded and wireless systems (ICNEWS)}.\hskip 1em plus 0.5em
  minus 0.4em\relax IEEE, 2018, pp. 1--5.

\bibitem{rtps25}
\BIBentryALTinterwordspacing
O.~M.~G. (OMG), ``The real-time publish-subscribe protocol dds interoperability
  wire protocol (ddsi-rtps) specification, version 2.5,'' 2021. [Online].
  Available: \url{https://www.omg.org/spec/DDSI-RTPS/2.5/About-DDSI-RTPS/}
\BIBentrySTDinterwordspacing

\bibitem{ros2securitytm}
\BIBentryALTinterwordspacing
T.~Moulard, J.~Hortala, X.~Perez, G.~Olalde, B.~Erice, O.~Olalde, and
  D.~Mayoral-Vilches, ``Ros 2 robotic systems threat model,'' 2019. [Online].
  Available: \url{http://design.ros2.org/articles/ros2_threat_model.html}
\BIBentrySTDinterwordspacing

\bibitem{ddssecurity11}
\BIBentryALTinterwordspacing
O.~M.~G. (OMG), ``Dds security, version 1.1,'' 2018. [Online]. Available:
  \url{https://www.omg.org/spec/DDS-SECURITY/1.1/About-DDS-SECURITY/}
\BIBentrySTDinterwordspacing

\bibitem{rivest1983cryptographic}
R.~L. Rivest, A.~Shamir, and L.~M. Adleman, ``Cryptographic communications
  system and method,'' Sep.~20 1983, uS Patent 4,405,829.

\bibitem{johnson2001elliptic}
D.~Johnson, A.~Menezes, and S.~Vanstone, ``The elliptic curve digital signature
  algorithm (ecdsa),'' \emph{International journal of information security},
  vol.~1, no.~1, pp. 36--63, 2001.

\bibitem{diffie1976new}
W.~Diffie and M.~Hellman, ``New directions in cryptography,'' \emph{IEEE
  transactions on Information Theory}, vol.~22, no.~6, pp. 644--654, 1976.

\bibitem{merkle1978secure}
R.~C. Merkle, ``Secure communications over insecure channels,''
  \emph{Communications of the ACM}, vol.~21, no.~4, pp. 294--299, 1978.

\bibitem{ros2securityintegration}
\BIBentryALTinterwordspacing
K.~Fazzari, ``Ros 2 dds-security integration,'' 2019. [Online]. Available:
  \url{http://design.ros2.org/articles/ros2_dds_security.html}
\BIBentrySTDinterwordspacing

\bibitem{ros2securityenclaves}
\BIBentryALTinterwordspacing
R.~White and M.~Arguedas, ``Ros 2 security enclaves,'' 2020. [Online].
  Available: \url{http://design.ros2.org/articles/ros2_dds_security.html}
\BIBentrySTDinterwordspacing

\bibitem{ros2securityaccesscontrol}
\BIBentryALTinterwordspacing
R.~White and K.~Fazzari, ``Ros 2 access control policies,'' 2019. [Online].
  Available:
  \url{http://design.ros2.org/articles/ros2_access_control_policies.html}
\BIBentrySTDinterwordspacing

\bibitem{multimachinedeployros2}
\BIBentryALTinterwordspacing
M.~Lanting, ``Added design document for remote and multi-machine launching,''
  2020. [Online]. Available: \url{https://github.com/ros2/design/pull/297}
\BIBentrySTDinterwordspacing

\bibitem{macenski2020marathon2}
S.~Macenski, F.~Martín, R.~White, and J.~G. Clavero, ``The marathon 2: A
  navigation system,'' in \emph{2020 IEEE/RSJ International Conference on
  Intelligent Robots and Systems (IROS)}, 2020, pp. 2718--2725.

\bibitem{macenski2021slamtoolbox}
\BIBentryALTinterwordspacing
S.~Macenski and I.~Jambrecic, ``Slam toolbox: Slam for the dynamic world,''
  \emph{Journal of Open Source Software}, vol.~6, no.~61, p. 2783, 2021.
  [Online]. Available: \url{https://doi.org/10.21105/joss.02783}
\BIBentrySTDinterwordspacing

\bibitem{schneider2003least}
F.~B. Schneider, ``Least privilege and more [computer security],'' \emph{IEEE
  Security \& Privacy}, vol.~1, no.~5, pp. 55--59, 2003.

\end{thebibliography}

\end{document}